\documentclass[pra,twocolumn,showpacs,floatfix]{revtex4-1}
\usepackage{natbib,times}
\usepackage{graphicx}
\usepackage{dcolumn}
\usepackage{amsmath}
\usepackage{amssymb}
\usepackage{color}

\def\beq{\begin{equation}}
\def\eeq{\end{equation}}

\def\om{\omega}
\def\a{\alpha}

\def\s{\sigma}
\def\D{\Delta}

\def\ad{a^\dagger}
\def\rd{{\rm{d}}}

\def\p{\phi}

\def\ra{\rightarrow}

\def\Zz{\mathbb{Z}}

\def\gb{\bar{g}}
\def\Db{\bar{\Delta}}
\def\Hp{{\cal H}_+}
\def\Hm{{\cal H}_-}
\def\Hpm{{\cal H}_\pm}

\def\pa{\partial}

\newcommand{\expec}[1]{\langle #1 \rangle}

\newcommand{\abs}[1]{\vert #1 \vert}
\newcommand{\ket}[1]{\vert #1 \rangle} 
 
\newcommand{\braket}[2]{\langle #1 \vert #2 \rangle}

\newcommand{\dg}{\dagger}

\newcommand{\ie}{i.e.\ }

\begin{document}
\title{Dynamical correlation functions and the quantum Rabi model}

\author{F. A. Wolf$^1$, F. Vallone$^2$, G. Romero$^2$, M. Kollar$^3$, E. Solano$^{2,4}$, and D. Braak$^1$}
\affiliation{$^{1}$ Experimental Physics VI, Center for Electronic Correlations and Magnetism, University of Augsburg, 86135 Augsburg, Germany}
\affiliation{$^{2}$ Departamento de Qu\'{i}mica F\'{i}sica, Universidad del Pa\'{i}s Vasco UPV/EHU, Apartado 644, 48080 Bilbao, Spain}
\affiliation{$^{3}$ Theoretical Physics III, Center for Electronic Correlations and Magnetism, University of Augsburg, 86135 Augsburg, Germany}
\affiliation{$^{4}$ IKERBASQUE, Basque Foundation for Science, Alameda Urquijo 36, 48011 Bilbao, Spain}
\date{\today}

\begin{abstract}
We study the quantum Rabi model within the framework of 
the analytical solution developed 
in Phys. Rev. Lett. {\bf 107}, 100401 (2011). 
In particular, through time-dependent correlation 
functions, we give a quantitative criterion 
for classifying two regions of the 
quantum Rabi model, involving the Jaynes-Cummings, 
the ultrastrong, and deep strong coupling regimes. 
In addition, we find a stationary qubit-field 
entangled basis that governs the whole dynamics as the 
coupling strength overcomes the mode frequency.
\end{abstract}

\maketitle

\section{Introduction}

The light-matter interaction has been subject of central interest since the early years of quantum mechanics. In $1936$, one of the first attempts to explain the results coming from experiments was the Rabi model, that describes the simplest dipole semiclassical interaction between light and matter~\cite{Rabi}, and is reduced to a pseudospin-$1/2$ system driven by a monochromatic classical radiation field. However, the advent of quantum technologies such as cavity QED~\cite{Walther2006,HarocheRaymondBook}, have allowed us to access the quantum regime of the radiation field, where the dynamical description is given by the celebrated Jaynes-Cummings model (JCM)~\cite{JaynesCummings}. This model predicts collapses and revivals of the population inversion, the appearance of Jaynes-Cummings doublets as a consequence of the excitation number conservation, and it has found a testbed in several hybrid setups such as trapped ions~\cite{Wineland2003}, quantum dots~\cite{hanson06}, and circuit quantum electrodynamics (circuit QED)~\cite{Wallraff2004,Blais2004,Chiorescu2004,Hofheinz2009,Schoelkopf2011,Esteve2011,Cleland2011,Houck2011,Houck2012,Abdumalikov2008}. 

Circuit QED has been growing both theoretically and experimentally, and their complex proposals such as the generation of multipartite entanglement~\cite{Wallraff09,Schoelkopf10,Majer07} rely on the fundamentals of the JCM and the Tavis-Cummings model~\cite{TavisCummings}. This can be justified because ratios between the coupling strength $g$ and the resonator frequency $\omega$ 
may grow from typical quantum optical values of $g/\omega\!\sim\!10^{-6}$ to circuit QED values of $g/\omega\!\sim\!10^{-2}$. Note that, in the latter, the rotating-wave approximation (RWA) can still be applied. Nowadays, two key experiments with superconducting circuits~\cite{Niemczyk10,Pol10} have made a significant improvement in the coupling strength, reaching values $g/\omega \sim 0.1$ in the so-called ultrastrong coupling (USC) regime~\cite{Ciuti05,Devoret2007,Bourassa2009}. In this case, the RWA is not longer valid and all dynamical and statical properties have to be explained through the quantum Rabi model (QRM) 
\beq
H_{R}=\hbar\Delta\s_z + \hbar\om\ad a + \hbar g\s_x(a+\ad),
\label{hamr}
\eeq
where $\sigma_z$ and $\sigma_x$ are Pauli matrices, $a (a^{\dag})$ is the annihilation (creation) operator, $\Delta\equiv\om_q/2$ is half of the qubit energy, $\omega$ is the resonator frequency, and $g$ stands for the coupling strength. In addition, a recent proposal considers the case where the coupling strength $g$ becomes comparable or larger than the mode frequency $\omega$, $g / \omega \gtrsim 1$, which is called deep strong coupling (DSC) regime~\cite{Casanova10}. In this case, the dynamics can be intuitively explained as photon number wavepackets propagating along a defined parity chain. Although the state-of-the-art in circuit QED does not provide this coupling strength yet, the main features of the DSC regime have been observed in an analog quantum simulation~\cite{Osellame2012,Ballester2012}.    

The QRM described by the Hamiltonian~(\ref{hamr}), has substantial differences as compared to the JCM, in fact only recently the properties of the QRM have been completely understood~\cite{Braak11,SolanoPhysics}. In the QRM, a discrete $\Zz_2$-symmetry replaces the continuous $U(1)$-symmetry of the JCM. Therefore the excitation number is no longer a conserved quantity and the Hilbert space splits into two infinite-dimensional invariant subspaces, the parity chains ~\cite{Casanova10}. 
Each eigenstate can be labeled with a  $\Zz_2$-quantum number, the parity $\hat{P} = \sigma_ze^{i\pi a^{\dag}a}$, taking values $\pm 1$. 

Furthermore, it is possible to give analytical expressions for these eigenstates as elements of the Bargmann space~\cite{Bargmann61}, that allows the well-defined computation of their norms and overlaps with eigenstates of the harmonic oscillator, without truncation of the Hilbert space~\cite{WolfBraak2012}.

The aim of this work is to present a new insight of the quantum Rabi model by stressing the use of dynamical correlation functions coming from the analytical solution obtained in Ref.~\cite{Braak11}. In this manner, we are able to explain relevant features such as the validity of the RWA in two well-defined regions, ranging from the JC regime to higher-coupling regimes of the quantum Rabi model. In addition, as the coupling strength $g$ enters into the DSC regime, we find that the true eigenstates of the system can be well approximated by the shifted oscillator basis in each invariant parity chain without the need for a more complicated  basis. This fact is supported by the calculation of the Wigner function of the eigenstates, whose unexpected fidelity can be understood in terms of the analytical form of the eigenfunctions~\cite{Braak11}, resembling closely to Fock states. In this context, we also find stationary Schr\"odinger cat-like states in the DSC regime. Finally, we present our concluding remarks.

\section{Dynamical classification}
\label{DC}
In this section, we shall use the exact dynamics of the quantum Rabi model to characterize two coupling regions that we may call {\it lower coupling region} and {\it higher coupling region}, respectively. The first region comprises the JC regime, where the RWA holds, together with the perturbative USC regime, where small deviations from the JC occur ($g /\om\lesssim 0.1$). The second identified region comprises a jump towards the higher-coupling regime, $g/\om \gtrsim 0.4$, and forms a precursor of the DSC regime ($g/\om > 1$). The intermediate region, that is for $0.1 /\om\lesssim g /\om\lesssim 0.4$, determines a kind of {\it dark zone} where no intuitive physics has been identified up to now.

We begin by studying a suitable dynamical observable, the time average photon number 
\beq
\bar{n}_{0} = \lim_{t\ra\infty}\frac{1}{t}\int_0^t\rd t' n_{0}(t') \, ,
\label{time-av}
\eeq
which measures the break of the $U(1)$-symmetry in the JCM. As we will show, this quantity exhibits significant features to characterize the regions mentioned before.

Since $H_R$ commutes with the parity operator $\hat{P}$, it is useful to investigate the dynamics within a fixed parity chain. Let $|\s\rangle$ with $\s=\pm1$ correspond to the states $|e\rangle$ and $|g\rangle$ of the two-level system, respectively.
Then the subspace $\Hpm$ with parity $\pm1$ is spanned by states $\{|\p_s\rangle\otimes|\pm\s\rangle,|\p_a\rangle\otimes|\mp\s\rangle\}$
where
$|\p_{(a),s}\rangle$ denotes the (anti-)symmetric part of $|\p\rangle$ which is an element of the Hilbert space ${\cal H}_b$ of the radiation mode. ${\cal H}_b$ is spanned by Fock states $\{|n\rangle\}$, which are (anti-)symmetric if $n$ is (odd) even.

There exists a transformation $F_\pm$ that maps the element $\ket{\phi}$ of ${\cal H}_b$ onto a parity eigenstate $\ket{\phi,\pm}$ that belongs to a parity chain
\beq
F_\pm|\p\rangle = |\p,\pm\rangle=
\p_s\otimes|\pm\s\rangle+\p_a\otimes|\mp\s\rangle.
\label{isoP}
\eeq
The dynamical quantities in each chain depend only on the initial distribution of photons, which fixes $|\p(t),\pm\rangle$ at time $t=0$. The state at later times follows as solution of the Schr\"odinger equation
\beq
i\pa_t|\p(t)\rangle=H_\pm|\p(t)\rangle,
\label{schP}
\eeq
where $H_\pm$ acts on functions in ${\cal H}_b$~\cite{WolfBraak2012}. The natural observable within the invariant subspaces is thus the photon number.

We consider first the time-dependent expectation value, $n_\p(t)=\langle\p(t),+|\ad a|\p(t),+\rangle$ for some initial $|\p(0)\rangle$ and positive parity. For $\D=0$ and an initial Fock state  $\ket{\p(0)}=|m\rangle=(a^\dg)^m/\sqrt{m!}|0\rangle$, the average photon number at time $t$ reads ($\gb=g/\om$)
\beq
n_m(t)=m+2\gb^2(1-\cos(\om t)),
\label{ndelta0}
\eeq
which entails that the difference $n_m(t)-m$ is independent of the initial photon number, always greater than zero and oscillates with the mode frequency $\om$. Because Fock states have definite reflection symmetry, the initial state in the full Hilbert space will be the product $|m\rangle\otimes|(-1)^m\rangle\in\Hp$.
\begin{figure}[t]
\includegraphics[width=0.44\textwidth]
{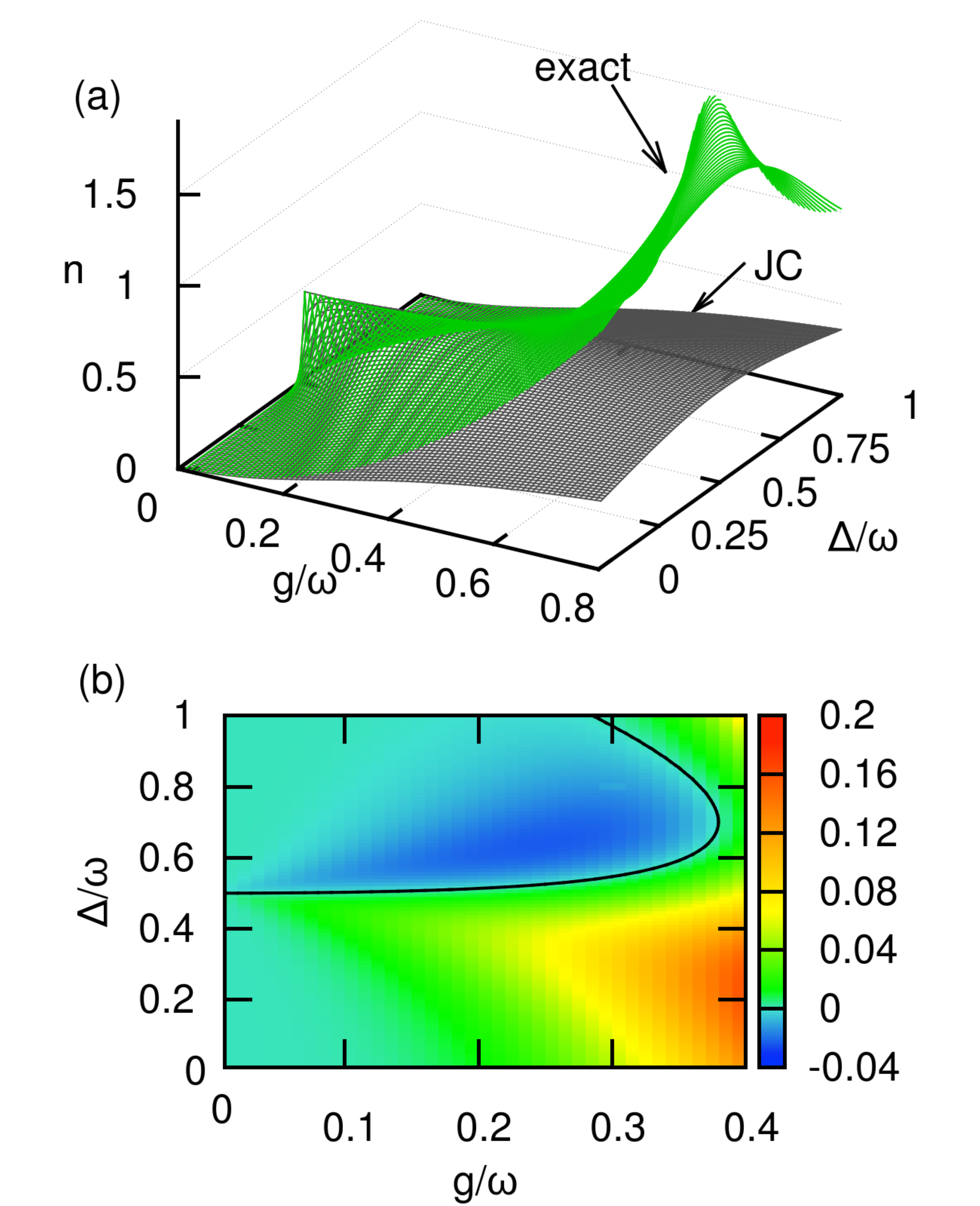}
\caption{(Color online)\ 
Panel (a) shows the time averaged photon number expectation value $\bar{n}_0$ -- the JC result of Eq.~\eqref{n0JC} and the exact result. Panel (b) shows the difference $\bar{n}_0^{\rm{exact}}-\bar{n}_0^{\rm{JC}}$. The black line shows where the JCM matches exactly the analytical solution.
}
\label{Fig1}
\end{figure}

Starting now from the photon vacuum $|0\rangle$ we consider values $\D\neq 0$. Then, the correlation function $n_0(t)$ depends on the initial state and shows a complicated oscillatory behavior. Still the time evolution can be globally characterized by the time average quantity defined in Eq.~(\ref{time-av}). Its dependence on $\Db\equiv\D/\om$ for fixed $\gb$ shows features that allow to discern 
a regime which corresponds to the validity of the RWA (the JC-regime) from a region where the
non-conservation of the excitation number in the QRM becomes relevant. 
As shown in Fig.~\ref{Fig1}(a), the Jaynes-Cummings model exhibits a sharp peak of $\bar{n}_0$ at $\Db=1/2$ with vanishing width as $g\ra 0$, and $\bar{n}_0\le 1/2$. The maximum value {1/2} is reached at resonance and corresponds to Rabi oscillations within the first JC-doublet, $\{|0e\rangle,|1g\rangle\}$. One photon is periodically exchanged between qubit and radiation field which contains 1/2 photon on average. For general values of $g$ and $\Delta$, the value of $n_0$ in the JCM reads 
\begin{align}\label{n0JC}
\bar{n}_0^{\rm{JC}}  & = \frac{2 g^2 d^2}{(g^2 + d^2)^2} \quad \\
\text{where}  \quad d & = \Delta-\tfrac{\om}{2}+\sqrt{(\Delta-\tfrac{\om}{2})^2 + g^2}. \nonumber
\end{align}
On the other hand, the exact result exhibits an unbounded $\bar{n}_0$ due to the broken $U(1)$-symmetry, together with a pronounced broadening of the resonance. Figure~\ref{Fig1}(b) shows the difference of the time-averaged photon number $\bar{n}_0$ calculated from the exact analytical solution and from the JCM. Notice that for $\Db = 1/2$, there is  good agreement between both quantities until a value $\bar{g}\sim 0.1$, establishing what we called before the lower-coupling region. This means that with regard to photon production, the RWA result is valid for couplings associated with the USC, 
at least exactly at $\Db=1/2$. For $\Db\neq 1/2$ the deviations set in much earlier.
Interestingly, for $\Db>1/2$, {\it less} photons are generated than expected from the RWA calculation.
This corresponds to a shift of the resonant $\Delta$ to values below $\om/2$.    
Finally, for couplings $\bar{g}\gtrsim 0.4$, what we called before the higher-coupling region, the resonance is completely gone and the photon production exceeds the RWA prediction for all $\Delta$, growing rapidly with the coupling. This marks the crossover to the deep strong coupling regime. Within the DSC, $g\gtrsim 1$, the qualitative behavior of the system changes again, accompanied by the stabilization of
Schr\"odinger cat-like
states
(see section \ref{dsc-regime}).

\section{Deep strong coupling features}

\begin{figure}[t]
\begin{center}
\includegraphics[width=0.44\textwidth]
{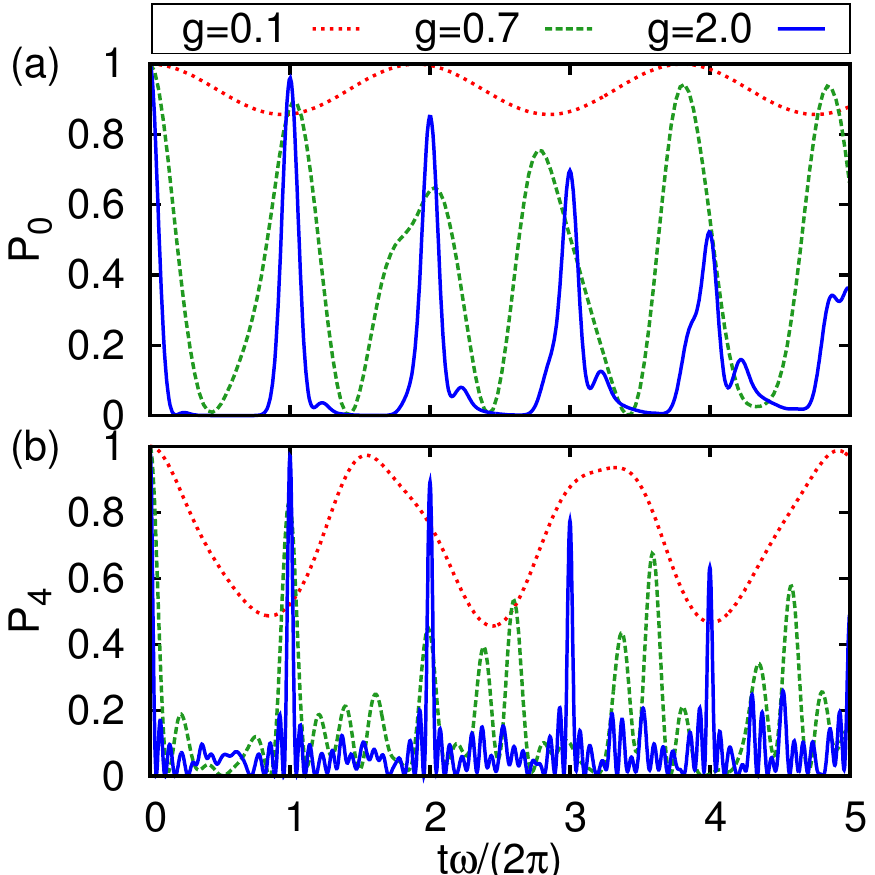}
\end{center} \vspace{-0.7cm}
\caption{(Color online)\ 
In (a) we show the revival probability $P_{{0}}(t)$ of state $\ket{\phi(0)}=\ket{0}\otimes\ket{e}$ and in (b) 
the revival probability $P_{{4}}(t)$ of state $\ket{\phi(0)}=\ket{4}\otimes\ket{e}$ for $\Db=0.25$ and three values of $\gb$. The result depicted is obtained by solving Eq.~\eqref{schP} for positive parity.}
\label{Fig2}
\end{figure}

We study the time evolution of the autocorrelation function $\langle\p(0),+|\p(t),+\rangle$ where we start with an initial Fock state with $\ket{\p(0)}=\ket{n}$. From this the revival probability is obtained as $P_n(t)=|\langle n ,+|\p(t),+\rangle|^2$. In Fig.~\ref{Fig2}, we observe the characteristic collapse and revivals of $P_n(t)$ in the deep strong coupling limit as in~\cite{Casanova10}. This feature is not present in the
USC but appears only for sufficiently strong g. Starting in the vacuum as in Fig.~\ref{Fig2}(a), the revivals are explained by the fact that the initial state $\ket{0}$ is a coherent state expressed in terms of the ground state of the quantum Rabi model $\ket{0}= e^{-\gb a^\dg} \ket{\psi_0}$ for $\Delta=0$. For large $g/\Delta$ the physical behavior of the $\Delta=0$ case is already present. That the physics of this limit is present already in a considerable distance from it is discussed in Sec.~\ref{secApproxBasis}. Collapse and revivals are also observed for higher initial Fock states as seen in Fig.~\ref{Fig2}(b). Then the revivals become sharper but have additional contributions from higher harmonics.

\begin{figure}[t]
\begin{center}
\includegraphics[width=0.46\textwidth]
{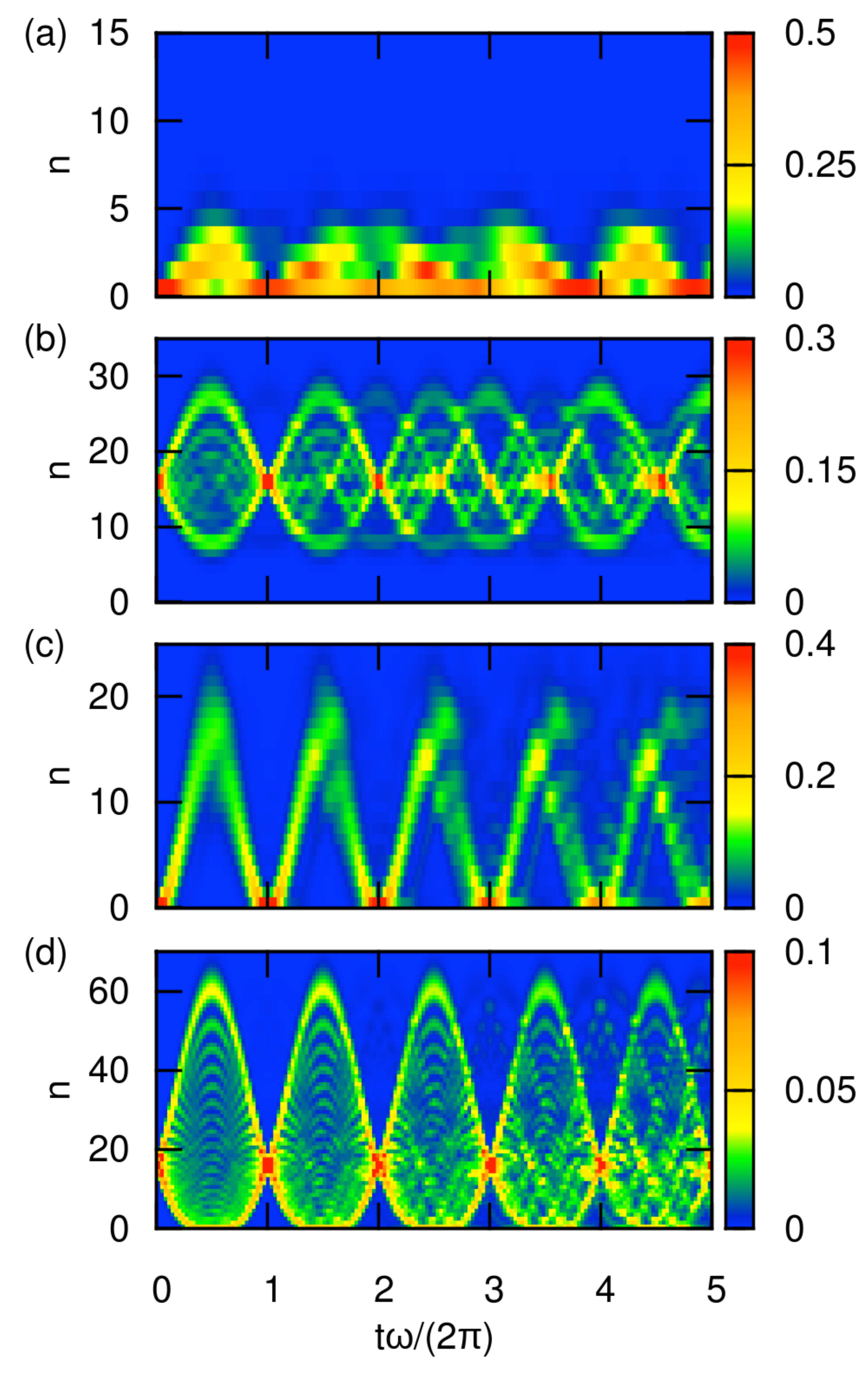}
\end{center} \vspace{-0.7cm}
\caption{(Color online)\ The figure shows the time-dependent photon number distribution $\abs{\braket{n}{\phi(t)}}^2$
for $\gb=0.7$ (a,b) and  $\gb=2.0$ (c,d) and different initial Fock states $\ket{\phi(0)}=\ket{0}\otimes\ket{e}$ (a,c) and $\ket{\phi(0)}=\ket{16}\otimes\ket{e}$ (b,d). $\Db=0.25$. The boundaries for the ``bouncing of photon number wave packets'' can be obtained in a simple way using Fig.~\ref{Fig4}(b).
}
\label{Fig3}
\end{figure}

In Fig.~\ref{Fig3}, we show the time evolution of the distribution obtained by projection on the complete Fock state basis, \ie the photon number distribution. Again the collapse and revival oscillations are visible in form of the red nodes of the interference patterns particularly pronounced in Fig.~\ref{Fig3}(d).
The second feature that is observed is the ``bouncing of photon number wave packets" \cite{Casanova10}.
Again the phenomenon can be understood in terms of the limit $\Delta=0$ where the eigen basis of the quantum Rabi model becomes the basis of a shifted oscillator. This is discussed in detail in the next section,  Sec.~\ref{secApproxBasis}. 

\subsection{Physics of the adiabatic approximation}
\label{secApproxBasis}

The natural starting point for an approximation of the full quantum Rabi model at large $g$ is the well known adiabatic approximation \cite{Schweber67}. The Hamiltonian to zeroth order for fixed parity reads
\beq
H_0=\om a^\dg a +g(a^\dg+a).
\label{hnull}
\eeq
The characteristic Rabi physics in the DSC can already be extracted by use of the eigenbasis $\{\ket{n;g}\}$ of this simple Hamiltonian. Any more sophisticated approximation to the eigenbasis improves the quality only marginally and does not reveal new physical behavior. For convenience we restrict ourselves to the positive parity subspace. The spectrum in $\Hp$ is given by the set of zeros  $x_n$ of the function~\cite{Braak11},
\beq
G_+(x)=\sum_{m=0}^\infty K_m(x)
\left[1-\frac{\D}{x-m\om}\right]\gb^m,
\label{sol}
\eeq  
as $E_n=x_n-g\gb$.
The functional form of $G_+(x)$ reads
\beq
G_+(x) = G_+^0(x) +\sum_{m=1}^\infty\frac{h^+_m(\D)}{x-m\om}
\label{GG0}
\eeq 
with $G_+^0(x)\sim (1-\D/x)e^{-\frac{x}{2\om}}$ and the $h^+_m(\D)$ vanish for $\D=0$.  $G_+^0(x)$ is of slow variation on the scale given by $\om$, therefore the location of the zeros of $G_+(x)$ is determined by the poles  at  $x=n\om, n=0,1,2\ldots$. This leads to an almost equidistant distribution of the $x_n$ for 
small $\D/g$:  The $n$-th root $x_n$ lies on the left or on the right of $n\om$ according to the sign of $h_n^+(\D)$. Because the $h_n^+(\D)$ grow with $\D$, we conclude that $x_{n+1}-x_{n}\approx \om$ for small $\D/g$. The dynamics of the model {\it for fixed parity} is encoded in this smooth distribution of eigenvalues together with  the fidelity of the adiabatic basis.

There is a natural association of the shifted oscillator basis $\{\ket{n;g }\}$ to the true eigenbasis $\{\ket{\psi_n}\}$ that can be inferred from the representation \cite{Braak11}, \beq\label{eqRepBarg}
\ket{\psi_n}=\sum_{m=0}^\infty K_m(x_n)\frac{\Delta \sqrt{m!}}{x_n-m\om} \ket{m;g} .
\eeq
As the quantum Rabi model approaches the adiabatic limit, the spectrum converges
to the spectrum of the shifted oscillator, \ie
\beq
x_n \rightarrow n\omega \quad \text{for} \quad g/\D \rightarrow \infty
\eeq
This limit is encoded in the representation \eqref{eqRepBarg} of the Rabi eigenstates by the divergence of 
\beq
(x_n-n\om)^{-1} \rightarrow \infty \quad \text{for} \quad g/\D \rightarrow \infty.
\eeq
This factor selects the $n$th shifted oscillator in the sum in Eq.~\eqref{eqRepBarg} for sufficiently large $g/\D$. We conclude that the Rabi eigenbasis in a fixed parity subspace is close to diagonal in the basis of shifted harmonic oscillators.

\begin{figure}[t]
\begin{center}
\includegraphics[width=0.46\textwidth]
{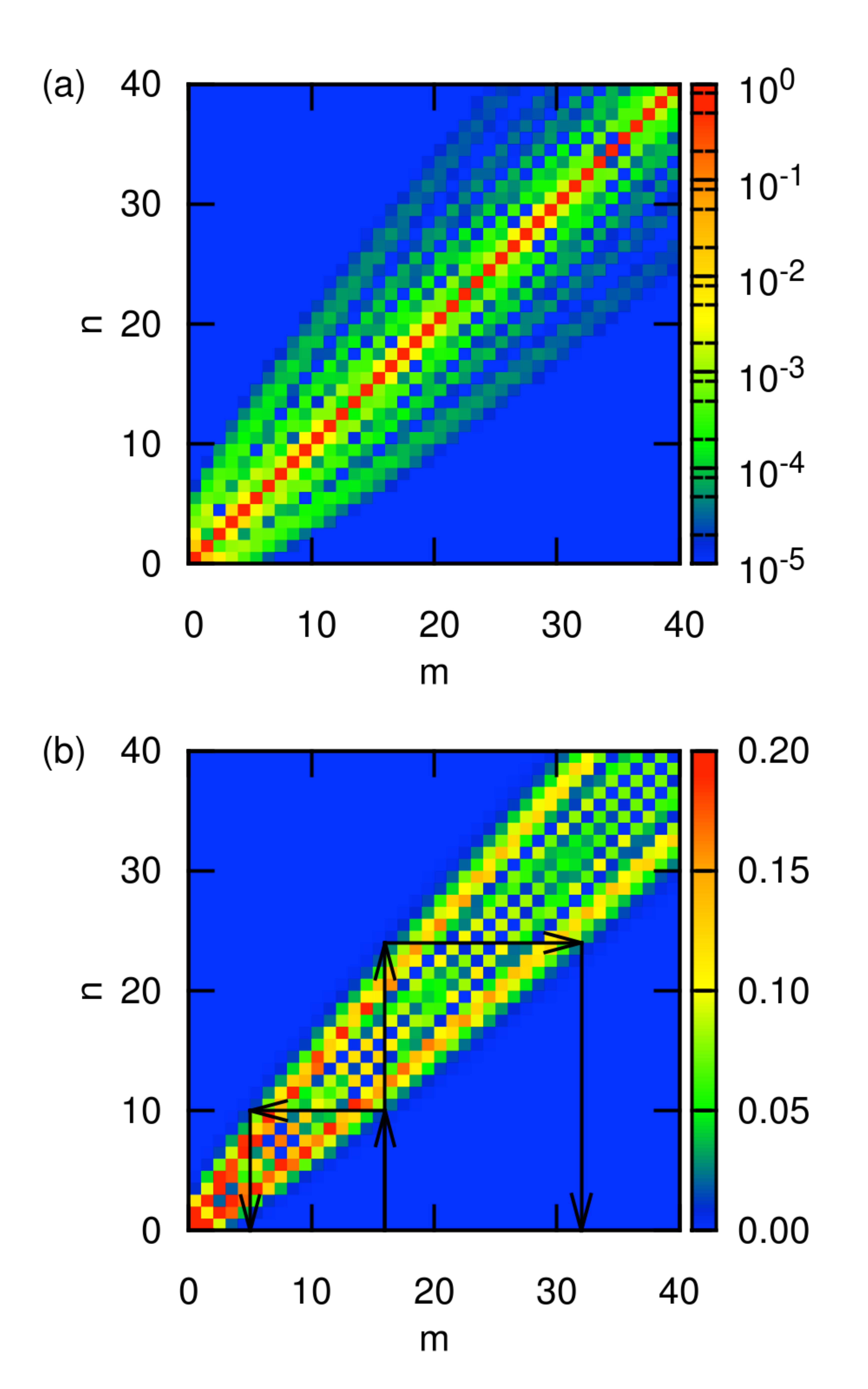}
\end{center} \vspace{-0.7cm}
\caption{(Color online)\  Density plot of the projection of eigenstates onto shifted oscillator states $\abs{ \braket{m;g}{\psi_n}}^2$ (a) and Fock states $\abs{ \braket{m;0}{\psi_n}}^2$ (b) for $\gb=0.7$ and $\Db=0.25$.
This is for positive parity, \ie $\psi_n$ is an eigenstate of $H_+$. The arrows in panel (b) show how to determine the boundaries for the ``bouncing photon number wave packets'' that are visible in Fig.~\ref{Fig3}.
}
\label{Fig4}
\end{figure}

This behavior is illustrated in Fig.~\ref{Fig4}(a) where we plot the contributions from the projections of the eigenstates on the shifted oscillators $\abs{\braket{m;g}{\psi^+_n}}^2$ on a logarithmic scale. The resulting matrix shows a rapid decrease of the off-diagonal elements. This reduction is already present for small values of $n$ and $m$ and becomes more pronounced for higher values.

In Fig.~\ref{Fig4}(b), by contrast, we plot the representation of the eigenstates in terms of Fock states. The figure displays a clear parabolic shape. This shape can be completely understood within the adiabatic limit. Considering the representation of the shifted oscillator states in terms of Fock states, a perfectly
symmetric parabola is seen. This is due to the fact that the Fock basis is transformed into the shifted oscillator basis by the operator $D(-\gb)=e^{-\gb a^\dg+\gb a}$ and the group property of $D$, $D(x)^{-1}=D(-x)$. Because the matrix element $D_{mn}(-x)$ reads $\braket{m;0}{x;n}=e^{-x^2/2}x^{m-n}\sqrt{m!/n!}L^{n-m}_m(x^2)$, where $L^{n-m}_m$ denotes a Laguerre Polynomial, it follows $|D_{nm}(x)|=|D_{nm}(-x)|$, furthermore, $D_{nm}(-x)=D^{-1}_{nm}(x)=D^\ast(x)_{mn}$ and therefore $|D_{nm}(x)|=|D_{mn}(x)|$.
The finite value of $\Db=0.25$ then leads to the slight asymmetry visible in the parabola of Fig.~\ref{Fig4}(b). Surprisingly, also Fig.~\ref{Fig4}(a) shows an almost symmetric shape which cannot be explained by the preceding argument.

\begin{figure}[t]
\begin{center}
\includegraphics[width=0.44\textwidth]
{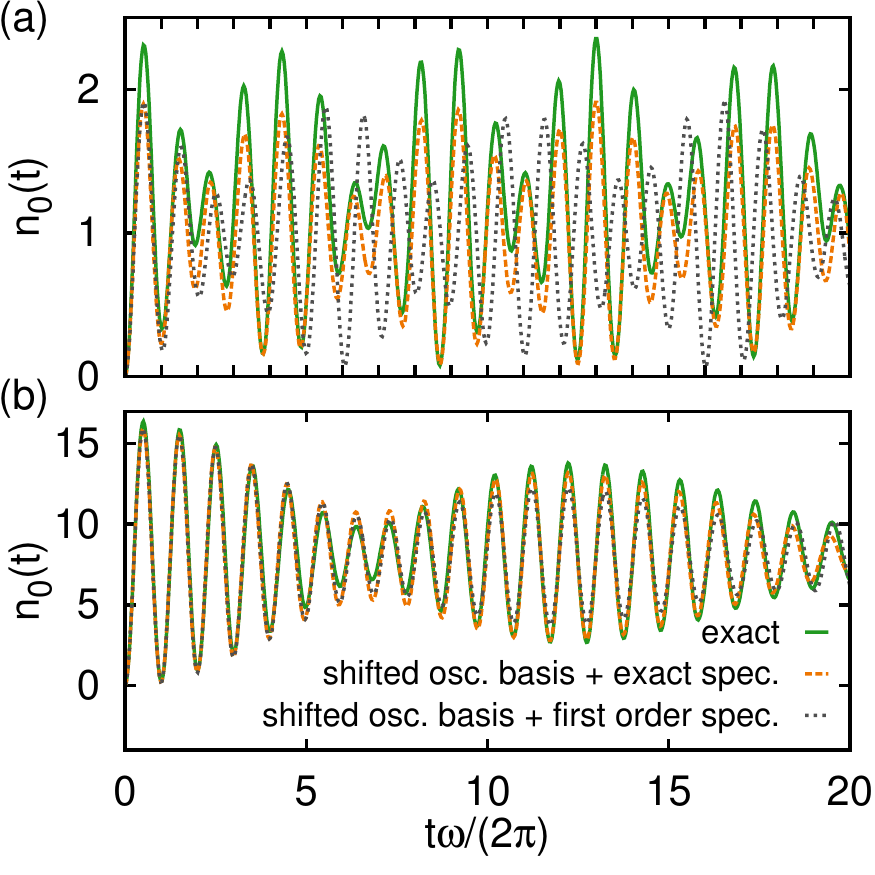}
\end{center} \vspace{-0.7cm}
\caption{(Color online)\ The time evolution of the photon number expectation value $n_0(t)$ for $\gb=0.7$ (a) and $\gb=2.0$ (b). $n_0(t)$ is evaluated as sum over the Photon distribution of Fig.~\ref{Fig3}(a) and (c):
$n_0(t) = \sum_m m \abs{\braket{\p_0(t)}{m}}^2$ where $\ket{\p_0(t)}= e^{-iH_+t} \ket{0} = \sum_n e^{-iE_n^+t} \ket{\psi_n}\braket{\psi_n}{0} $. (i) solid lines depict numerically exact results calculated with this expression, (ii) dashed lines are calculated with $\ket{\psi_n}$ replaced by $\ket{n,g}$ and (iii) dotted lines are calculated with the additional replacement of the spectrum by its first order corrected perturbation theory value used in Eq.~\eqref{distance2}. For $\gb=0.7$, the adiabatic approximation allows a good description of the time evolution only if the exact spectrum is employed as seen in panel (a). For $\gb=2.0$, the adiabatic approximation is highly precise also if the approximated spectrum is used.
}
\label{Fig5}
\end{figure}

The effect of bouncing photon number wave packets mentioned earlier and shown in Fig.~\ref{Fig3} can be described solely by the presence of the parabola in Fig.~\ref{Fig4}(b). To illustrate the determination of the boundaries for the bouncing phenomenon in Fig.~\ref{Fig3}(b), we have added the arrows in Fig.~\ref{Fig4}(b). The upwards pointing arrow indicates the initial state $\ket{16}$. The horizontal arrows then lead to the minimum and maximum photon numbers contained in the eigenstates.

In Fig.~\ref{Fig5}, we present results for the time evolution of the quantum Rabi model calculated using the adiabatic approximation. Fig.~\ref{Fig5}(b) shows that for $g/\Delta=8$ the adiabatic approximation gives a detailed picture of the full time dependence. Fig.~\ref{Fig5}(a) by contrast shows that for moderate values of the coupling, here $g/\Delta\simeq2.8$, the full adiabatic approximation, \ie approximated spectrum and approximated basis, completely misses a physical description of the dynamics. This insufficiency of the full adiabatic approximation is mainly  due to deviations in the spectrum which  yield wrong phases. 
By contrast, the approximated basis combined with the exact spectrum leads to a qualitatively correct description of the dynamics.

\begin{figure}[t]
\begin{center}
\includegraphics[width=0.46\textwidth]
{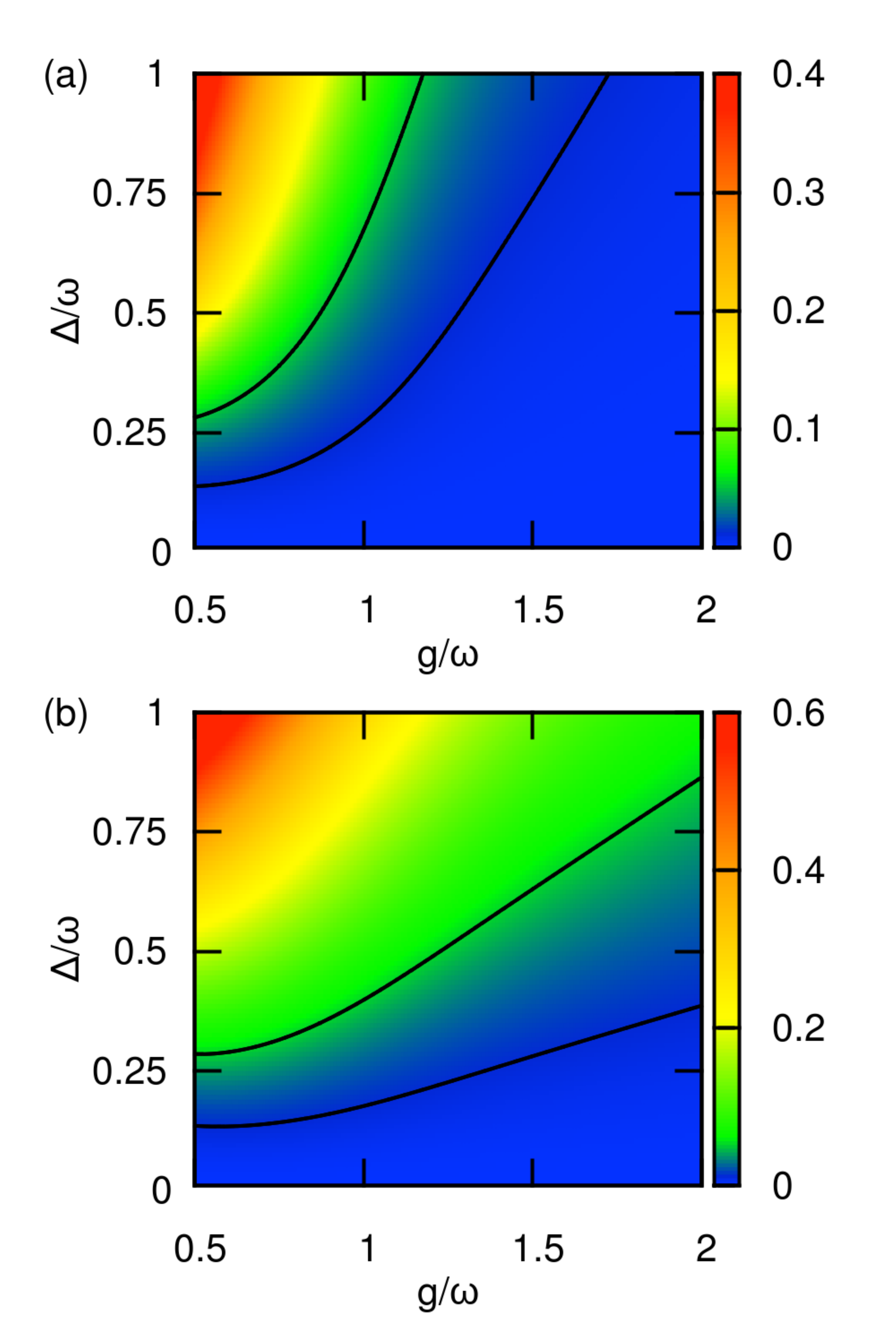}
\end{center} \vspace{-0.7cm}
\caption{(Color online)\ The distance $D_0(\gb,\Db)$ of the shifted oscillator ground state from the ground state of the quantum Rabi model in the positive parity subspace is shown in panel (a). The corresponding difference of the eigen energies $D^{E}_0(\gb,\Db)$ in panel (b). The contour lines in the plots are drawn at values of $0.01$ and $0.05$. Comparing the slope of the contour lines between panel (a) and (b) shows that the quality of the basis increases faster than the quality of the spectrum. 
}
\label{Fig6}
\end{figure}

To further illustrate the point that the adiabatic basis contains already all of the physics even far away from 
the limit $g/\D\rightarrow\infty$, while the adiabatic spectrum leads to unphysical results, we define a distance of the state $\ket{n,g}$ of the approximate basis from the true eigenstate $\ket{\psi_n}$ by
\beq \label{distance}
D_n(g,\D)= 1- \abs{\braket{n;g}{\psi_n}}^2.
\eeq
A distance of the first order corrected eigen energy from the true eigen energy is given by the difference 
\beq \label{distance2}
D^{E}_n(g,\D)= \abs{\om n - g\gb + \Delta (-1)^n \braket{n,g}{n,-g} - E_n}.
\eeq
These functions are shown in Fig.~\ref{Fig6} for $n=0$. While their magnitude cannot be compared
directly, their dependence on $g$ and $\Delta$ allows to conclude the following. The slope of the contour lines is much steeper in case of the basis (Fig.~\ref{Fig6}(a)) compared to the spectrum (Fig.~\ref{Fig6}(b)).
This means that the quality of the approximate basis increases much faster with $g/\D$ than that of the first order spectrum. This is surprising, as the basis is only a zeroth order approximation while the spectrum is approximated to first order.

It is possible to improve the adiabatic basis in a systematic fashion \cite{Feranchuk96,Pan10} or use the RWA on top of the adiabatic approximation \cite{Irish07}. There are other approaches which reduce the approximative diagonalization of $H_R$ in $\Hpm$ to  diagonalization of finite matrices \cite{Pereverzev06,Liu09}. All these techniques are equivalent to a break-up of $\Hpm$ into a set of invariant subspaces with finite dimension. This means that a continuous symmetry is superimposed on the quantum Rabi Hamiltonian, leading to an effective model of Jaynes-Cummings type \cite{Braak11}. In the quantum Rabi model, such an additional (hidden) symmetry is not even present in an approximate sense, because otherwise Fig.~\ref{Fig4}(a) would display a block-diagonal pattern. Instead, the off-diagonal components of the true eigenstates in terms of shifted oscillator states are distributed rather smoothly on both sides of the central diagonal being at the same time smaller by at least two orders of magnitude. It can be concluded that no self-consistent reduction to finite-dimensional invariant subspaces \cite{Pereverzev06,Irish07} is possible which improves the adiabatic basis without creating a strong (and unphysical) symmetry. Most of the physics in the DSC regime (in fact, already for $\gb\gtrsim 0.7$)
is captured by the simple adiabatic basis.  

Besides these physical considerations, we point out the technical fact that the spectrum is much easier to compute than the basis. Combining the exact spectrum with the simple basis allows us to predict correctly dynamical quantities of the QRM in the blue sketched parameter region of Fig.~\ref{Fig6}(a),  as demonstrated in Fig.~\ref{Fig5}.

\subsection{Eigenstate Wigner functions}

\begin{figure}[t]
\begin{center}
\includegraphics[width=0.52\textwidth]
{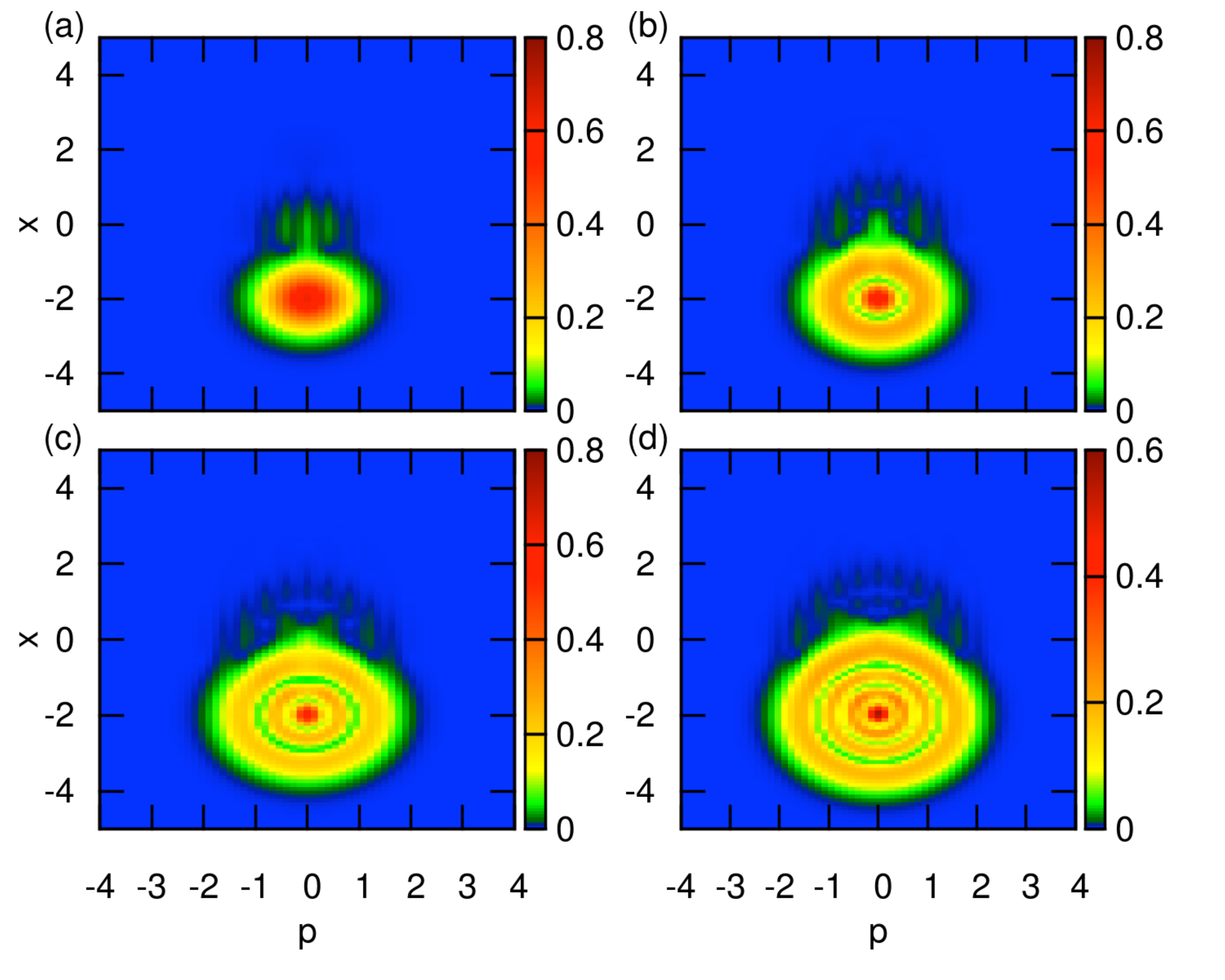}
\end{center} \vspace{-0.7cm}
\caption{(Color online)\  Wigner functions of eigenstates in the parity chain $\cal{H}_{-}$, and considering the DSC regime. (a) Grouns state, (b) first excited, (c) second excited, and (d) third excited states. We can see that they resemble the Wigner functions  of Fock states $|0\rangle$, $|1\rangle$, $|2\rangle$, $|3\rangle$, respectively, but displaced from the center as expected from the shifted oscillator basis.}
\label{Fig7}
\end{figure}

The above discussion in the DSC regime led us to find the shifted basis as a good approximation to the real one. The feature can be further confirmed by analyzing the Wigner functions associated to each eigenstate in both parity chains, $\mathcal{H}_{\pm}$. Fig.~\ref{Fig7} shows the Wigner functions corresponding to the first eigenstates in the parity chain $\mathcal{H}_-$. We see that these functions resemble  the Fock states $|0\rangle$, $|1\rangle$, $|2\rangle$, and $|3\rangle$, though shifted from the center, as expected from the shifted oscillator basis. Notice  the appearance of some additional interference pattern coming  from other Fock states which contribute to the whole solution, see Fig.~\ref{Fig4}(a). It is noteworthy to mention that exactly the same behavior can be observed in the other
 parity chain $\mathcal{H}_{+}$. Both chains behave very similar regarding the
form of the eigenvectors, the ground state in $\Hm$ looks almost the same as the first exited state
in $\Hp$.

\subsection{Stationary Schr\"odinger cat-like states}
\label{dsc-regime}
\begin{figure}[t]
\begin{center}
\includegraphics[width=0.46\textwidth]
{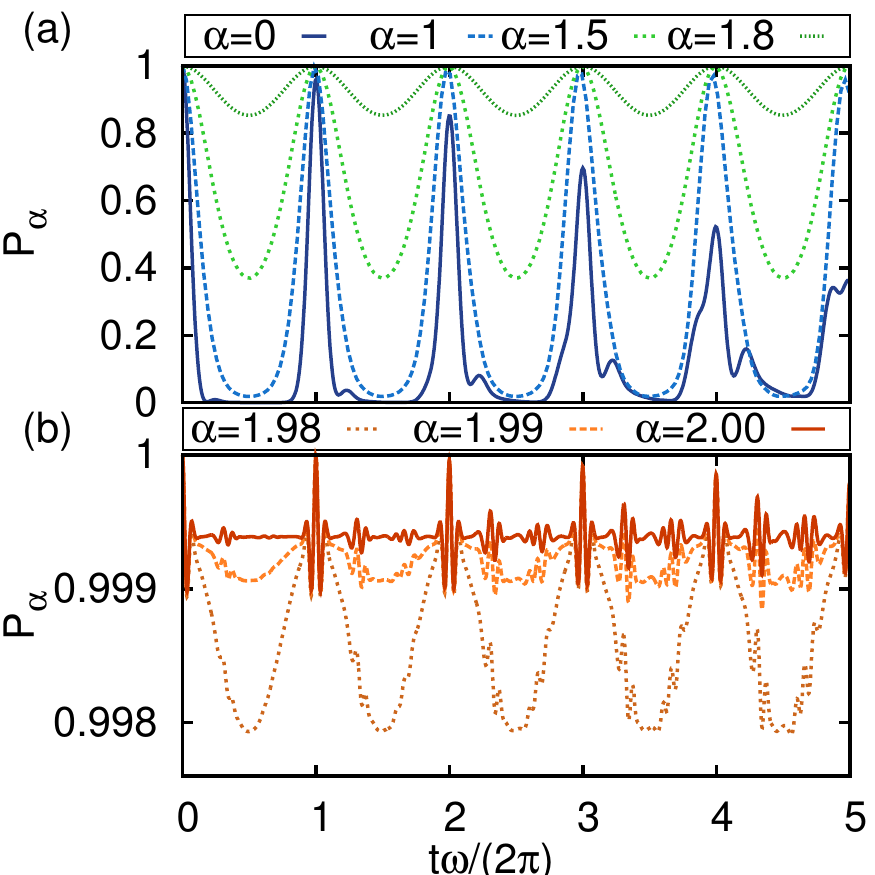}
\end{center} \vspace{-0.7cm}
\caption{(Color online)\  Revival probability $P_{\alpha}(t)$ of coherent states $e^{\a\ad}|0\rangle$ for different values of $\alpha$ and $\Db=0.25$. If $\alpha=\gb$, the time evolution of the coherent state remains almost constant, 
$P_{\alpha=\gb}(t) \equiv P_{C_+}(t) \sim 1$ for all times. This is to be compared with the time evolution depicted in Fig.~\ref{Fig1}(b) of the initial Fock state $\ket{4}$ that yields the same photon number $\expec{\hat n}=4$ as $\ket{C_+}=e^{2\ad}|0\rangle$.
}
\label{Fig8}
\end{figure}

Fig.~\ref{Fig6}(a) shows that the ground state of $H_+$ at strong coupling is well approximated by $\ket{0;g}= e^{-\gb^2/2-\gb\ad}|0\rangle$. If the system is prepared in this state, it will only weakly depend on time.
Using the transformation $F_\pm$ from Eq.~\eqref{isoP}, the state $\ket{0;g}$ is mapped to the following states in the original basis with fixed parity $\pm1$:
\beq
\begin{array}{cc}
|C_+\rangle=&
e^{-\frac{\gb^2}{2}}\big(\cosh(\gb a^\dg)\ket{0}\otimes|e\rangle-\sinh(\gb a^\dg)\ket{0}\otimes|g\rangle\big) 
\\
|C_-\rangle=&
e^{-\frac{\gb^2}{2}}\big(\cosh(\gb a^\dg)\ket{0}\otimes|g\rangle-\sinh(\gb a^\dg)\ket{0}\otimes|e\rangle\big)
\end{array}
\label{cat-states}
\eeq
These Schr\"odinger cat-like states contain qubit-field entanglement~\cite{Brune92}, where the terms $\sinh(\gb a^\dg)\ket{0}$ and $\cosh(\gb a^\dg)\ket{0}$ are (anti-)symmetric superpositions of the semiclassical states $e^{-\gb\ad}|0\rangle$ and $e^{\gb\ad}|0\rangle$~\cite{Ashhab10,Ciuti11}. Indeed, as the average photon number $\langle n\rangle$ in  $e^{\pm\gb\ad}|0\rangle$ is $\gb^2$, an experimentally realizable value of $\langle n\rangle=4$ corresponds to a DSC value $\gb=2$. These states behave almost like eigenstates regarding expectation values of observables, e.g. the revival probability $P_{C_\pm}(t)$. As seen in Fig.~\ref{Fig8}, $P_{C_\pm}(t) \sim 1$ remains almost constant for $\a=\gb$ and exhibits only small high-frequency variations, but not the collapse-and-revival behavior with frequency $\om$ as the Fock
states in Fig.~\ref{Fig1}.  

Fig.~\ref{Fig8}(b) shows that, for $\a=\gb$, the oscillations with $\om$ are completely gone and all fluctuations occur on time scales much shorter than the oscillator period $2\pi/\om$. Averaged over such a short time, $P_{C_\pm}(t)$  remains constant and behaves effectively a conserved quantity although the corresponding state is not an eigenstate. The phase variable of the protected coherent states depends directly on the coupling $\gb$ and confines the effect to the DSC regime if the coherent states are required to contain more than two photons on average.

\section{Conclusions}

We have studied the dynamical properties of the quantum Rabi model. By using the recently developed analytical solutions~\cite{Braak11}, we  obtained qualitative features of the spectrum and the eigenstates in two parts of the QRM: a lower and a higher coupling region, respectively. It turns out that it is mandatory to pay attention to the sole and important  $\Zz_2$-symmetry of the QRM, which leads to the separation of the Hilbert space into two invariant subspaces, the parity chains~\cite{Casanova10}. Within each chain, the dynamics appears to be rather simple. Especially in the deep strong coupling regime, $\gb \gtrsim 1$, we find a quite regular periodic behavior of the photon number distribution which can be traced back to the almost equidistant separation of energy levels. This is no longer true if the initial state has no definite parity. Under such circumstances, e.g. if the initial state is a product of a coherent state and a qubit eigenstate, the time development may be more complicated, as both chains interfere~\cite{WolfBraak2012}.

We have identified a simple observable, the average photon generation from the vacuum at positive parity, to discern the convenient lower-coupling and higher-coupling regions. This quantity shows a remarkable sharp peak just for $\D=\om/2$, which demonstrates in a direct way the resonant enhancement of coupling between qubit and cavity mode for small $g$, which was the original motivation to introduce the RWA in the first place~\cite{JaynesCummings}. This determines the lower-coupling region. For stronger couplings, the resonance gets broader until the peak vanishes around $g/\om\sim 0.4$, establishing the higher-coupling region. Here, the  average generated photon number becomes larger than the JC limit of 1/2 due to the counter-rotating terms in (\ref{hamr}), which break the conservation law of the JCM.

As early as for $g/\om\gtrsim0.7$, the characteristic features of the deep strong coupling regime begin to manifest. The dynamics for fixed parity are dominated by the adiabatic basis and the almost equidistant spectrum. Interestingly, the nontrivial effects, which separate the QRM in this regime  from the simple adiabatic limit, can be incorporated by using the exact spectrum together with the adiabatic basis. In fact, the exact eigenstates are very close to their adiabatic approximants whereas deviations in the eigenvalues lead to phase differences which become apparent after longer times (Fig.~\ref{Fig5}). The fidelity of the adiabatic basis is even more visible in the Wigner representation, which demonstrates the similarity of parity eigenstates with Fock states in the deep strong coupling regime (Fig.~\ref{Fig7}). 

Finally, the DSC allows for a special class of states which are not exact eigenstates but lead to expectation values fluctuating with small amplitudes on very short time scales. In this sense, they form stationary Schr\"odinger cat-like states, unaffected by the system interaction. 

\section*{ACKNOWLEDGMENTS} 
We acknowledge funding from the Deutsche Forschungsgemeinschaft through TRR 80, Spanish MICINN Juan de la Cierva, FIS2009-12773-C02-01, Basque Government IT472-10, UPV/EHU UFI 11/55, SOLID, CCQED, and PROMISCE European projects.

\end{document}